\documentclass[conference,a4paper]{IEEEtran}

\usepackage{xcolor}
\usepackage{balance}
\ifCLASSINFOpdf
  \usepackage[pdftex]{graphicx}
\else
  \usepackage[dvips]{graphicx}
\fi
\usepackage{cite}
\usepackage{epstopdf}
\usepackage{multirow}
\usepackage{siunitx}
\usepackage{amsmath}
\usepackage{siunitx}
\usepackage{pgfplots}
\pgfplotsset{compat=newest}
\usepackage{nicefrac}
\newlength\figureheight
\newlength\figurewidth
\newcommand{\figref}{Fig.~\ref}

\usepackage{comment}
\usepackage{dblfloatfix} 
\ifCLASSOPTIONcompsoc
 \usepackage[caption=false,font=normalsize,labelfont=sf,textfont=sf]{subfig}
\else
 \usepackage[caption=false,font=footnotesize]{subfig}
\fi
\usepackage[absolute,overlay]{textpos} 

\begin{document}

\title{Flexible Multimode-Based Beamforming MIMO Antenna}

\author{\IEEEauthorblockN{
Abel Zandamela\IEEEauthorrefmark{1},   
Nicola Marchetti\IEEEauthorrefmark{1},   
Adam Narbudowicz\IEEEauthorrefmark{1}    
}                                     
\IEEEauthorblockA{\IEEEauthorrefmark{1}%
CONNECT Centre, Trinity College Dublin, The University of Dublin, Dublin, Ireland.\\ \{zandamea, nicola.marchetti, narbudoa\}@tcd.ie}
}

\maketitle

\begin{abstract}
This work proposes compact, flexible Multiple Input Multiple Output (MIMO) antennas. The design principle is based on the excitation of different orthogonal radiating modes within the same antenna volume. Via phase control of the excited modes, beamforming is demonstrated in azimuth and elevation planes using single-layered structures. For flexibility, the antennas are designed using Polydimethylsiloxane (PDMS) as the substrate. Numerical results demonstrate that isolation better than 23 dB is realized in all investigated antennas under different bend configurations. Moreover, the proposed technique demonstrates an antenna with unidirectional beamsteering across the entire elevation plane, and a second design realizes a bi-directional beamsteering in the horizontal plane. Overall, the results highlight the potential of multimode-based beamforming for flexible MIMO antennas in Internet of Things (IoT) systems. 
\end{abstract}

\vskip0.5\baselineskip

\section{Introduction}
The popularity of smart wearable electronics increasingly places high demands of novel small-size and multi-functional antenna systems. This is because, in addition to the tight size specifications, wearables need to be flexible and have minimal performance deterioration when operating in proximity to the human body \cite{Iurii2021, Islam2020, Qadri2020}. Furthermore, to support advanced wireless applications, e.g. on-/off-body centric communications, MIMO, localization via Angle of Arrival (AoA) estimation, the antenna needs to generate multiple patterns \cite{Simorangkir2017, Alqadami2016, IoTMAG2022, Obeidat2021}, which further increases the overall design complexity.  

Flexible multi-functional antennas have been investigated, e.g. in \cite{Simorangkir2017, Alqadami2016, Li2018, Quoc2022, GuoPing2022}. In \cite{Quoc2022}, a planar inverted-F antenna and varactors are used for a dual-band frequency-reconfigurable wearable textile antenna. Shorting pins
and arc-shaped slots are used in \cite{Simorangkir2017} to realize a single-layer dual-band dual-mode antenna for on-/off-body centric communications. The study in \cite{GuoPing2022}, uses two radiators and shorting pins to design a double-layered dual-band dual-mode antenna with improved frequency tuning capabilities. However, the aforementioned designs, either allow for frequency reconfiguration or can only generate two discrete radiation patterns (omnidirectional/broadside patterns) in different frequency bands. Even though such performance is applicable for body-centric communications, the integration of switches limits their use for applications requiring the simultaneous use of multiple patterns, e.g. MIMO or AoA-based localization.      

MIMO flexible antennas have been studied, e.g. in \cite{Alqadami2016,Li2018,Peng2019,Yang2022}. A multi-layered MIMO antenna comprising rectangular microstrip subarrays with $0.28\lambda$ spacing (where $\lambda$ is the wavelength at the antenna center operating frequency) is proposed in \cite{Alqadami2016}.~In \cite{Li2018}, a single-layer dual-polarized dipole textile MIMO antenna is proposed for wearable applications. The work in \cite{Yang2022} investigates a triple-band dual-polarized MIMO belt strap antenna for wearables in healthcare. The above-mentioned works, present significant advances for MIMO systems while also allowing for flexible and compact structures; nevertheless, drawbacks still exist as some designs are multi-layered and bulky due to inter-spacing between the elements in the array, or have limited beamforming capabilities.    

\begin{figure}[!b]
\centering
{\includegraphics[width=1.0\columnwidth]{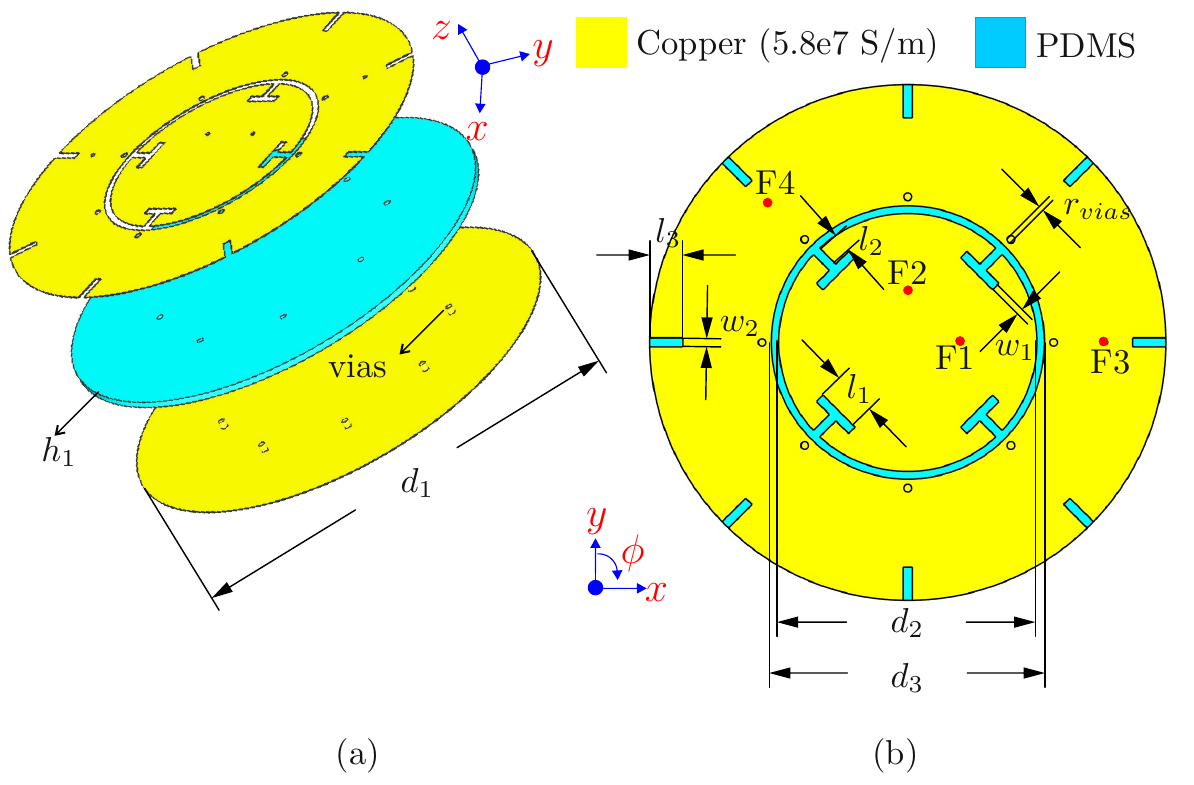}}
\caption{Antenna 1: (a) exploded view and (b) top view.}
\label{fig:system_setupBroad}
\end{figure}

\begin{figure}[!b]
\centering
{\includegraphics[width=1.0\columnwidth]{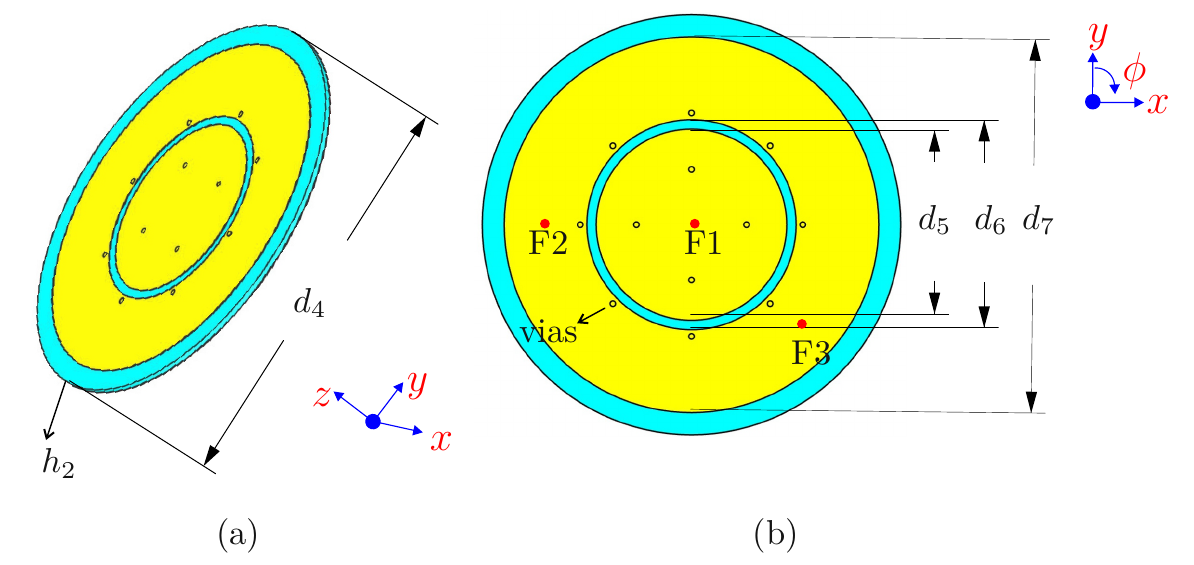}} \caption{Antenna 2: (a) perspective view and (b) top view.}
\label{fig:system_setupAzi}
\end{figure}

In this work, we propose a multimode-based beamforming concept for advanced beamsteering in MIMO flexible antennas. The principle is based on the excitation of different orthogonal modes, using patch and annular ring antennas printed in a single substrate layer. First, a four-port antenna is proposed to realize beamforming in the elevation plane; next, a three-port design is proposed for beamsteering across the entire horizontal plane. The performance is analyzed in free space without bend and further validated with bends along the antenna $x-$ and $y-$axes. While different techniques are used to design flexible substrates, e.g. liquid crystal polymers \cite{Vyas2007, Iurii2021}, Parylene-N \cite{Sharifi2009}; here, flexible Polydimethylsiloxane (PDMS) \cite{Wang2012} substrates are used to realize the proposed MIMO antennas. The PDMS material is chosen because it has been shown to provide high flexibility, low dielectric loss, thermal stability, and water resistance, while also allowing good tunability of its dielectric permittivity \cite{Mata2005, Wang2012}.

\section{Proposed Antennas Configuration}

The proposed antennas are shown in \figref{fig:system_setupBroad} and \figref{fig:system_setupAzi}. Each design comprises a central circular patch, a shorted-ring structure, and a Polydimethylsiloxane (PDMS) layer.

\subsection{Antenna 1 - Elevation Plane Beamforming} \figref{fig:system_setupBroad} shows Antenna 1, designed to operate around $\SI{5.7}{GHz}$, with total dimensions: $\SI{34}{mm}\times \SI{34}{mm}\times \SI{0.62}{mm}$ or $0.64\lambda \times 0.64\lambda \times 0.012\lambda$. The metallization layer has $\SI{0.06}{mm}$ thickness, and the PDMS layer is $h_1=\SI{0.5}{mm}$ thick. The central patch (\figref{fig:system_setupBroad}b), has diameter $d_2 = \SI{17}{mm}$, and incorporates four T-shaped slots for miniaturization; the slots are rotated by $\ang{90}$ with respect to the disk center. The T-shaped horizontal slot has length $l_1=\SI{3}{mm}$ and width $w_1=\SI{0.6}{mm}$; the vertical slot length is $l_2=\SI{1.49}{mm}$ and width $w_2=\SI{0.6}{mm}$. The patch is fed using two ports located at F1 ($\SI{3.5}{mm}$,~$\SI{0}{mm}$) and F2 ($\SI{0}{mm},~ \SI{3.5}{mm}$), exciting two orthogonal broadside modes. The second structure is a concentric ring with inner diameter $d_3 =\SI{18}{mm}$, and outer diameter $d_1 = \SI{34}{mm}$. The ring is shorted using 8 vias of radius $r_{vias} = \SI{0.25}{mm}$, located at $\SI{0.6}{mm}$ from the inner edge and rotated by $\ang{45}$ with respect to the disk center. Eight slots of length $l_2 = \SI{2.2}{mm}$, and width $w_2=\SI{0.6}{mm}$, are used for frequency tuning. The shorted-ring is fed using two ports located at F3 ($\SI{13}{mm}$,~$\SI{0}{mm}$) and F4 ($\SI{-9.2}{mm}$,~$\SI{9.2}{mm}$), to excite two orthogonal phase-varying TM$_{21}$ modes.

\subsection{Antenna 2 - Horizontal Plane Beamforming}
\figref{fig:system_setupAzi} shows Antenna 2, a 3-port MIMO design proposed for beamforming in the horizontal plane ($xy-$plane). The design operates around $\SI{5.76}{GHz}$, and the total size is $\SI{37}{mm}\times \SI{37}{mm}\times \SI{0.62}{mm}$ or $0.71\lambda \times 0.71\lambda \times 0.012\lambda$. The metallization and PDMS layers have similar thicknesses as the Antenna 1, respectively $\SI{0.06}{mm}$ and $h_2=\SI{0.5}{mm}$. Its central patch (diameter $d_5=\SI{17.3}{mm}$) is shorted using four vias of radius $r_{vias} = \SI{0.25}{mm}$, located at $\SI{5}{mm}$ from the disc center and rotated by $\ang{90}$. The patch is center-fed F1 ($\SI{0}{mm}$,~$\SI{0}{mm}$) and excites a monopole-like mode with a constant phase around the entire horizontal plane. The second radiator is a shorted-ring structure exciting two orthogonal omnidirectional phase-varying modes. The ring outer diameter is $d_7=\SI{34}{mm}$, and the inner diameter is $d_6=\SI{19}{mm}$. The ring vias are located $\SI{0.6}{mm}$ from the inner edges and are rotated by $\ang{45}$ with respect to the disk center. The feeding points are F2 ($\SI{-13.5}{mm}$,~$\SI{0}{mm}$), and F3 ($\SI{9.5}{mm}$,~$\SI{-9.5}{mm}$), exciting two orthogonal TM$_{21}$ modes.

\section{Simulations and Flexibility Analysis}

\begin{figure}[b]
\centering
{\includegraphics[clip, trim=0cm 21cm 5.8cm 0cm,width=.8\columnwidth]{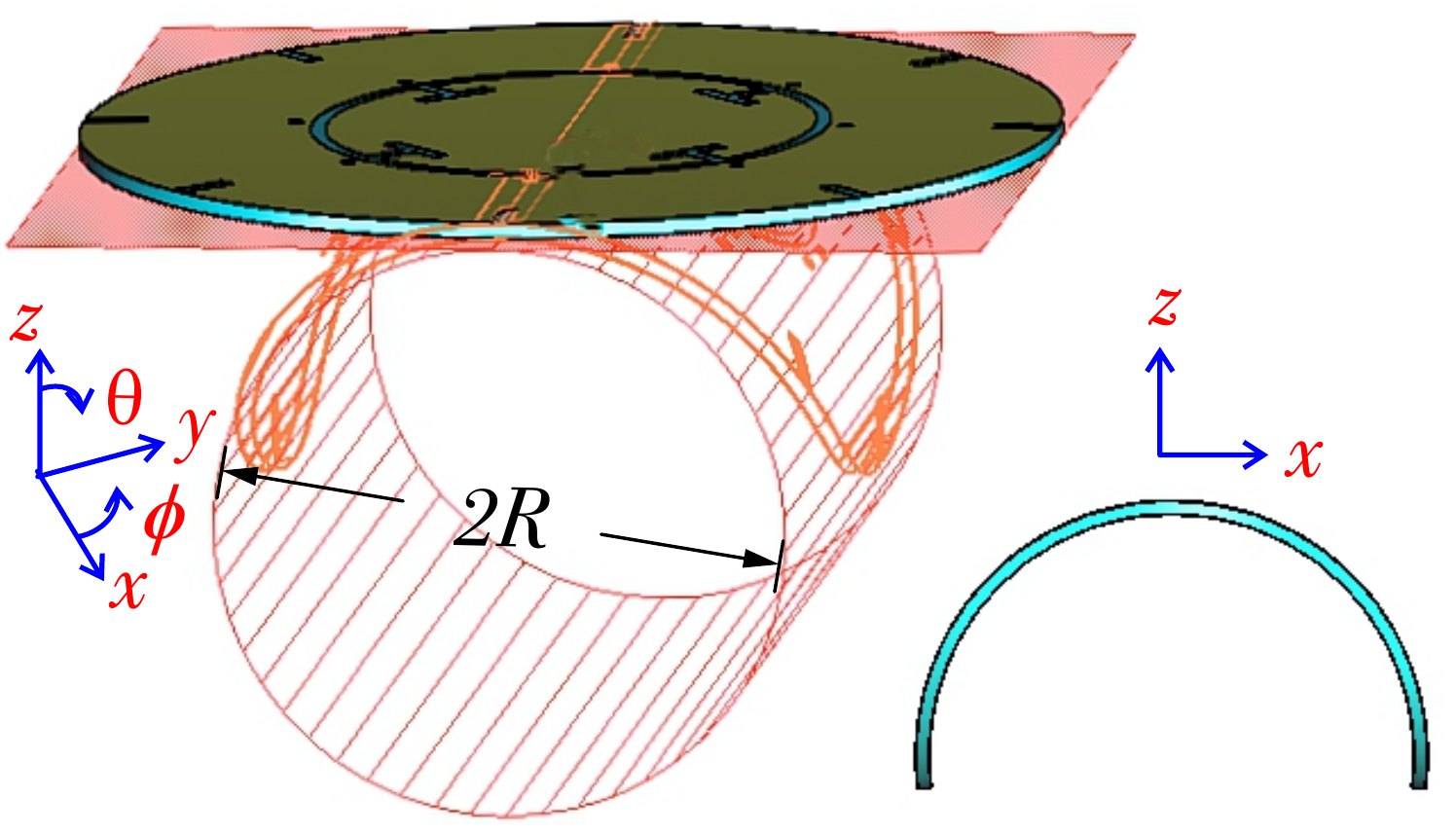}} \caption{Proposed bending procedure for the flexibility analysis: (left) cylindrical bends along the $x-$axis realized using radius $R=\SI{10}{mm}$; and (right) front view showing the obtained configuration for the $x-$axis bend.}
\label{fig:bentView}
\end{figure}

This Section discusses the simulation results of the antennas presented in Section II. The results are obtained from a 3D Finite Element Method (FEM) full-wave solver (CST).

\subsection{Broadside Beamforming Antenna}

\begin{figure}[!t]
{\includegraphics[clip, trim=1.8cm 5.5cm 11cm 5.4cm, width=1.0\columnwidth]{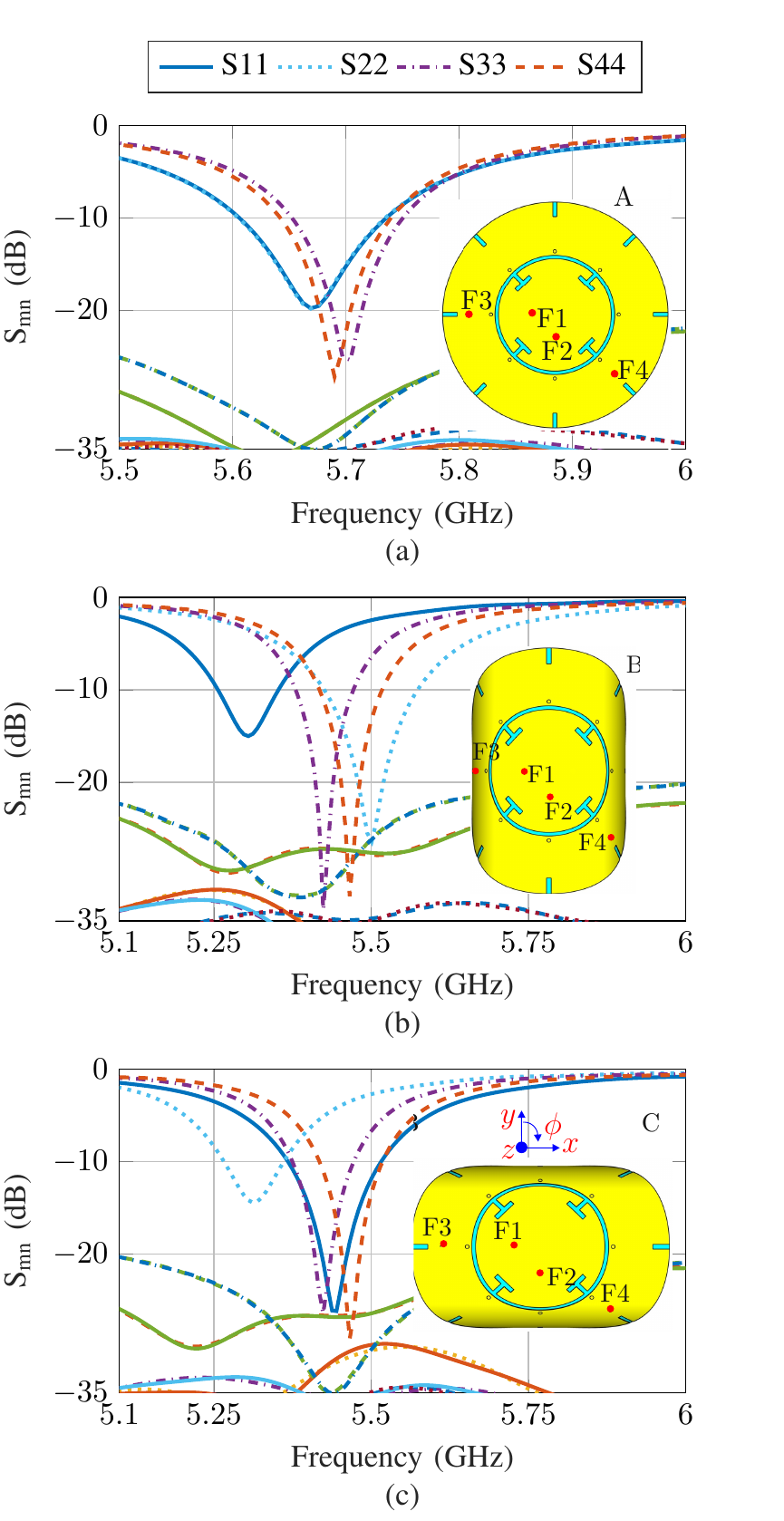}}
\caption{Antenna 1 simulated S-parameters: (a) antenna without any bends; (b) with bends along the $x-$axis; and (c) bends along the $y-$axis. Note that the isolation is not shown for clarity, but it is at least better than $\SI{26}{dB}$.}
\label{fig:SparaBroad}
\end{figure}

The S-parameters of Antenna 1 (\figref{fig:system_setupBroad}) are shown in \figref{fig:SparaBroad}a. Owing to the use of orthogonal radiating modes, isolation  $>\SI{33}{dB}$ is realized between all four antenna ports. The $\SI{-10}{dB}$ impedance bandwidth, overlapping between all the ports is around $\SI{83}{MHz}$. 

To investigate the flexibility of the proposed design, cylindrical bends with radius ($R=\SI{10}{mm}$) are introduced in the antenna around the $x-$ and $y-$axes. The bending procedure is outlined in \figref{fig:bentView}. Only the procedure for the bend along the $x-$axis is shown for brevity. To realize the $y-$axis bend, it will only require positioning the cylinder along the same axis. The simulated S-parameters are shown in \figref{fig:SparaBroad}b and \figref{fig:SparaBroad}c. The results show that bending the structure results in a frequency shift towards lower bands. The shift can be explained by the increase in the current path for the bent structures, especially around the bent regions of the patch and ring antennas. 

In detail, \figref{fig:SparaBroad}b shows that for the central patch, the $x-$axis bend results in a relatively larger frequency shift for F1 ($S_{11}$ curve), which may be explained by the fact that the feed is much closer to the bent regions. This is different for F2 ($S_{22}$ curve), which is on the $x=0$ line, and for the bend along the $x-$axis, the effects on the current path are symmetrical when viewed from this line, i.e. the bends are symmetric from both $-x$ and $+x$ directions. This leads to a minor shift of $S_{22}$ compared to $S_{11}$. For the shorted-ring, the $S_{44}$ has a smaller shift compared to $S_{33}$, as the F3 feed is located in the region with a larger bend, resulting in an increased surface current path compared to F4. \figref{fig:SparaBroad}b shows that due to the relatively larger frequency shift for F1, its radiated field will have a weak contribution to the steered pattern. In contrast, F2, F3, and F4 still show an overlapping bandwidth of $\SI{51}{MHz}$ around $\SI{5.45}{GHz}$. As demonstrated in the following subsection, beamsteering covering the entire elevation plane is still feasible under these conditions.

\figref{fig:SparaBroad}c shows the S-parameters for the $y-$axis bends. Opposite to the bend along $x-$axis, in this configuration, the relatively larger frequency shift is seen for $S_{22}$. This shift is also generated due to the feed location of port F2, which is away from the $y=0$ line, and much closer to the bent regions, resulting in a much larger shift when compared to F1. Because F1 is located on the $y=0$ line, the effects on the current path from this feeding point are symmetrical when viewed from ($-y$ and $+y$). For this configuration an overlapping bandwidth of $\SI{55}{MHz}$ is also seen around $\SI{5.45}{GHz}$ for F1, F3 and F4. This frequency range is subsequently studied for beamforming, and similarly to what was observed for the $x-$axis bend, the port with the largest shift will have a weaker contribution to the generated pattern and is not included in the beamforming analysis.

\begin{figure}[!b]
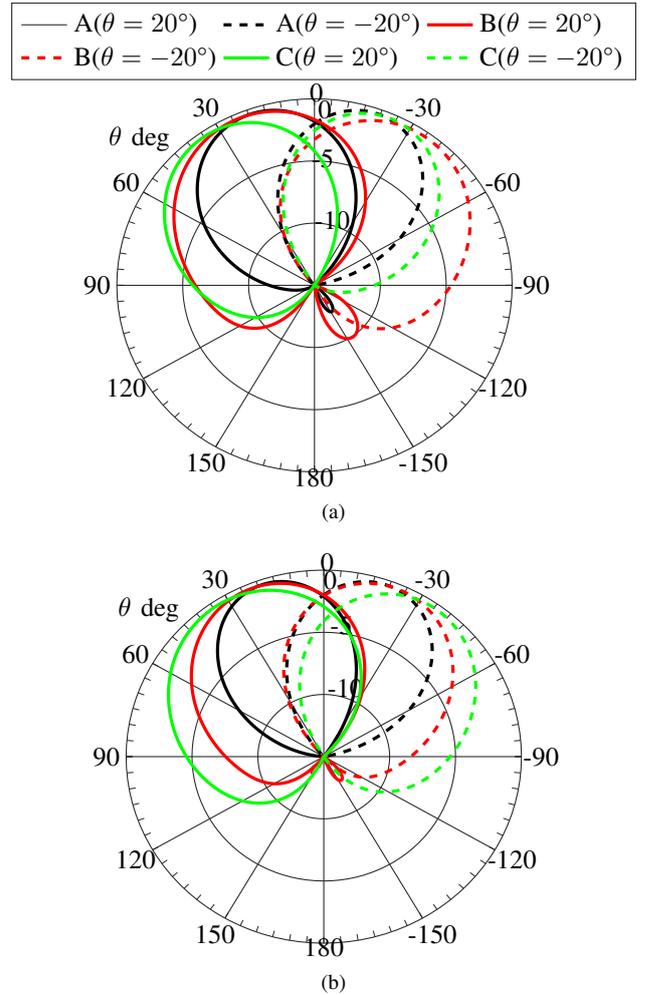

\centering
\subfloat[]{
\hspace{-0.75cm}
\setlength\figureheight{0.3\textwidth}
\setlength\figurewidth{0.4\textwidth}
\input{scan_xz_cst}}
\hfill
\subfloat[]{
\hspace{-0.75cm}
\setlength\figureheight{0.3\textwidth}
\setlength\figurewidth{0.4\textwidth}
\input{scan_yz_cst}}
\caption{Antenna 1 beamsteering: (a) $xz-$plane; and (b) $yz-$plane.}
\label{fig:ElevSteer}
\end{figure}

\subsection{Elevation Plane Beamforming}

To show the potential of the proposed beamforming principle for flexible MIMO antennas, the design capability to steer the beam along the elevation plane is shown in \figref{fig:ElevSteer}. The performance is tested at $\SI{5.7}{GHz}$ for the case without bend (denoted A). For the bends along the $x-$ and $y-$axes, respectively denoted by B and C, the beamsteering is tested for the overlapping region of F1/F2 and F3 and F4, which is around $\SI{5.45}{GHz}$, as shown in \figref{fig:SparaBroad}.

\begin{figure}[!t]
\centering
\hspace{.25cm}
{\includegraphics[clip, trim=1.8cm 5.5cm 11cm 5.2cm, width=.9\columnwidth]{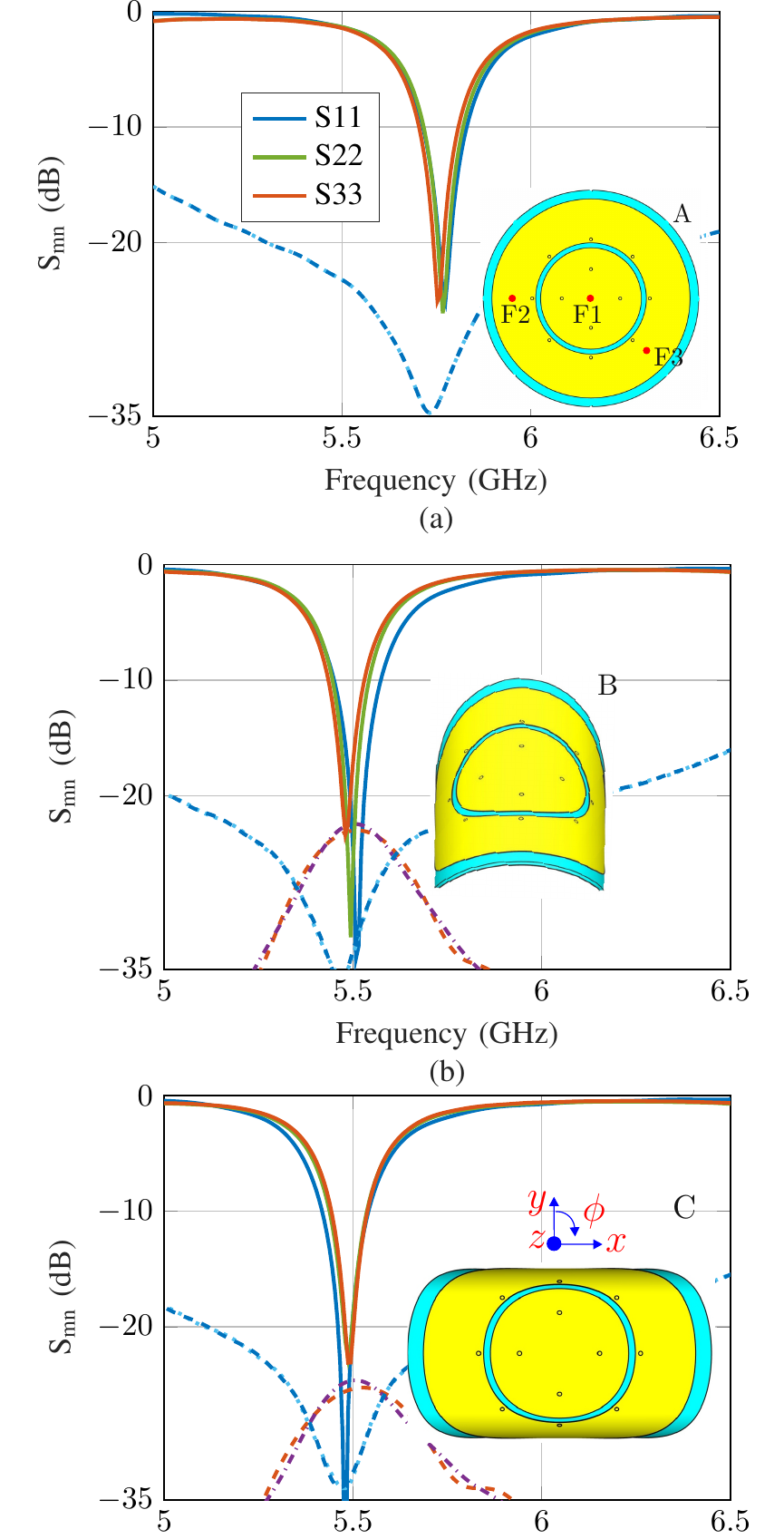}}
\caption{Antenna 2 simulated S-parameters: (a) antenna with no bends; (b) $x-$axis bends and (c) bends along $y-$axis.}
\label{fig:SparaAzi}
\end{figure}

\figref{fig:ElevSteer}a shows the beamsteering in the $xz-$plane for two different directions: $\theta=\ang{20}$ and $\theta=\ang{-20}$, shown in solid and dashed lines, respectively. It can be seen that all three configurations are capable of realizing beamsteering across the $xz-$plane. For the configuration A, the steering towards $\theta=\ang{20}$ is obtained by only activating ports F2 and F4, with a phase shift of $\ang{70}$ for the broadside radiating port (F2); to direct the beam towards $\theta=\ang{-20}$, ports F1 (with $\ang{60}$ phase shift) and F3 are used. For the antenna with bend along $x-$axis (B), $\theta=\ang{20}$ is obtained from F2 (with $\ang{150}$ phase shift) and F4; the $\theta=\ang{-20}$ is generated by combining F2 and F4 with no additional phase shift. Note that the configuration B results in a much larger half-power beamwidth of around $\ang{115}$ (see \figref{fig:ElevSteer}a) compared to configurations A (beamwidth around $\ang{75}$), and C ($\ang{85}$). This is mostly because the bends around the $x-$axis lie within the steering plane ($xz-$plane). The beamsteering for configuration C (shown in green curves) is obtained by combining F1 (with $\ang{-120}$ phase shift) and F3 for $\theta=\ang{20}$; for $\ang{-20}$ F3 and F1 are fed with an additional $\ang{180}$ phase shift, i.e. $\ang{-20} + \ang{180}=\ang{160}$ is used. 

\figref{fig:ElevSteer}b shows the beamsteering around the $yz-$plane for all the three configurations. It is seen that a good beamsteering characteristic is also achieved across this plane for all three cases. For configuration A, the steering towards $\theta=\ang{20}$ is realized by F1 (with $\ang{70}$ phase shift) and F4, while for the $\theta=\ang{-20}$ direction the ports F2 (with $\ang{-100}$ phase shift) and F3 are used. Because the antenna in case B, has a frequency shift for F1, this port is not used for beamsteering in this configuration. By combining F2 (with $\ang{60}$ phase shift) and F3, the beam can be steered towards $\theta=\ang{20}$; the $\theta=\ang{-20}$ direction can be covered by using the same ports but with an  additional $\ang{180}$ phase shift, giving $\ang{240}$. Lastly, because F2 is shifted for the configuration C the beamsteering is realized from F1 and F4; $\ang{-10}$ phase shift in F1 directs the beam towards $\theta=\ang{20}$ and adding $\ang{180}$ phase shift to the initial setting gives $\ang{170}$, which directs the beam towards $\theta=\ang{-20}$.

\subsection{Horizontal Plane Radiating Design}

The simulated S-parameters of Antenna 2 are shown in \figref{fig:SparaAzi}a. Its center frequency is around $\SI{5.77}{GHz}$, where the isolation between all the antenna ports is $>\SI{33}{dB}$, and the overlapping bandwidth is around $\SI{94}{MHz}$. \figref{fig:SparaAzi}b and \figref{fig:SparaAzi}c show the S-parameters of the antenna with bends along the $x-$ and $y-$axes, respectively. In both cases, a downward frequency shift is observed due to the bends. Both structures have a center frequency around $\SI{5.49}{GHz}$. Because F1 is centrally located, the effects of the bend are symmetrical when viewed from either ($-x$, $+x$) for case B or ($-y$, $+y$) for C. This allows for an overlapping bandwidth of $\SI{80}{MHz}$ in both configurations; however, the isolation changes to around $\SI{23}{dB}$. 

\subsection{Horizontal Plane Beamforming}

\begin{figure}[!b]
\centering
\subfloat[]{
\setlength\figureheight{0.3\textwidth}
\setlength\figurewidth{0.4\textwidth}
\input{Bidir/bid_0deg}}
\hfill
\subfloat[]{
\setlength\figureheight{0.3\textwidth}
\setlength\figurewidth{0.4\textwidth}
\input{Bidir/bid_45deg}}
\hfill
\subfloat[]{
\setlength\figureheight{0.3\textwidth}
\setlength\figurewidth{0.4\textwidth}
\input{Bidir/bid_90deg}}
\caption{Antenna 2 beamsteering: (a) $\phi=\ang{0}$; (b) $\phi=\ang{45}$, and (c) $\phi=\ang{90}$.}
\label{fig:AziSteer}
\end{figure}

The beamsteering in the horizontal plane is achieved by introducing phase shifts on the ports of the shorted ring (F2 and F3, which excite the phase-varying TM$_{21}$ modes). The phase shifts are computed with respect to the desired direction to create a constructive interference (see \cite{AbelAEU}, and \cite{EuCAP2022} for more details). The beamsteering performance is tested for three different directions: $\phi=\ang{0}, \ang{45}$, and $\ang{90}$, shown respectively in \figref{fig:AziSteer}a, b and c. \figref{fig:AziSteer}a shows that the three configurations are capable of steering the beam towards the desired $\phi=\ang{0}$ direction; however, due to the small number of the excited omnidirectional modes, a bi-directional pattern is observed. This ambiguity can be eliminated by using additional phase-varying modes \cite{AbelAEU}. Note that the configuration C shows wider beamwidth of $\ang{110}$ compared to A ($\ang{83}$) and B ($\ang{66}$); this is because the $\phi=\ang{0}$ direction is orthogonal to the bend in configuration C ($y-$axis bend). In contrast, the configuration B has a narrower beamwidth as the bends are in the same plane as the desired direction. For $\phi=\ang{90}$ direction, the performance is shown in \figref{fig:AziSteer}c, where all the configurations also have their main beams pointing at the desired angle. In this direction, the wider beamwidth is seen for case B, while a narrower one is seen for case C (as the bends lie around $y-$axis). \figref{fig:AziSteer}b shows the performance around $\phi=\ang{45}$. It can be seen that the patterns become more tilted if the desired direction is located away from the center of the bends and does not lie in a plane orthogonal to the directions of the bends. However, the configuration B still has the main beams pointing around the $\phi=\ang{30}$ and the $\ang{30} + \ang{180}$ ambiguity, with a narrow beamwidth performance; while the configuration C has the main beams in the orthogonal plane. Overall, our results demonstrate the potential multimode excitation to realize compact beamsteering flexible MIMO antennas for emerging smart wireless systems, like wearables and healthcare.

\section*{Acknowledgment}
This publication has emanated from research conducted with the financial support of Science Foundation Ireland under Grant number $18$/SIRG/$5612$.

\balance


\begin{thebibliography}{1}

\bibitem{Iurii2021}
I. Cherukhin, S.-P. Gao, and Y. Guo, ``Fully Flexible Polymer-Based Microwave Devices: Materials, Fabrication Technique, and Application to Transmission Lines,'' \emph{IEEE Trans. Antennas Propag.}, vol.69, no.12, pp.8763-8777, 2021.

\bibitem{Islam2020}
G. M. N. Islam, A. Ali, and S. Collie, ``Textile sensors for wearable applications: A comprehensive review,'' \emph{Cellulose}, vol. 27, no. 11, pp. 6103–6131, Jul. 2020.

\bibitem{Qadri2020}
Y.~A.~Qadri et al., ``The Future of Healthcare Internet of Things: A Survey of Emerging Technologies,'' \emph{IEEE Comm. Surveys \& Tutorials}, vol. 22, no. 2, pp. 1121-1167, Secondquarter 2020.

\bibitem{Simorangkir2017}
R. B. Simorangkir, et al., ``Dual-Band Dual-Mode Textile Antenna on PDMS Substrate for Body-Centric Communications,'' \emph{IEEE Antennas Wirel. Propag. Lett.}, vol. 16, pp. 677-680, 2017.

\bibitem{Alqadami2016}
A. S. M. Alqadami, et al., ``Assessment of PDMS technology in a MIMO antenna array,'' \emph{IEEE Antennas Wireless Propag. Lett.}, vol. 15, pp. 1939–1942, 2016.

\bibitem{Obeidat2021}
H.~Obeidat, et al.,~``A Review of
Indoor Localization Techniques and Wireless Technologies,'' \emph{~Wirel.~Pers.
Comm.},~vol.~117,~no.~4,~pp. 1–39, Feb. 2021.

\bibitem{IoTMAG2022}
A. Zandamela, A. Chiumento, N. Marchetti and A. Narbudowicz, ``Angle of Arrival Estimation via Small IoT Devices: Miniaturized Arrays vs. MIMO Antennas,'' \emph{IEEE Internet of Things Mag.,} vol. 5, no. 2, pp. 146-152, June 2022.

\bibitem{Quoc2022}
Q. H. Dang, S. J. Chen, and C. Fumeaux, ``Dual-Band Frequency-Reconfigurable Flexible Wearable Textile Antenna,'' \emph{2022 IEEE AP-S/URSI}, pp.397-398, 2022.

\bibitem{GuoPing2022}
G. P. Gao et al., ``Dual-Mode Patch Antenna With Capacitive Coupling Structure for On-/Off-Body Applications,'' \emph{IEEE Antennas Wirel. Propag. Lett.}, vol.21, no.8, pp.1512-1516, 2022.

\bibitem{Li2018}
H. Li, S. Sun, B. Wang and F. Wu, ``Design of Compact Single-Layer Textile MIMO Antenna for Wearable Applications,'' \emph{IEEE Trans. Antennas Propag.}, vol. 66, no. 6, pp. 3136-3141, June 2018.

\bibitem{Peng2019}
J. P. Xu et al., ``Flexible Wearable Antenna based on MIMO Technology,'' \emph{2019 IEEE MTT-S Int. Wirel. Symp. (IWS)}, pp.1-3, 2019.

\bibitem{Yang2022}
S. Yang et al., ``Flexible Tri-Band Dual-Polarized MIMO Belt Strap Antenna Toward Wearable Applications in Intelligent Internet of Medical Things,'' \emph{IEEE Trans. Antennas Propag.}, vol.70, no.1, pp.197-208, 2022.

\bibitem{Sharifi2009}
H. Sharifi et al., ``Characterization of Parylene-N as flexible substrate and passivation layer for microwave and millimeter-wave integrated circuits,'' \emph{IEEE Trans. Antennas Propag.}, vol. 32, no. 1, pp. 84–92,
Feb. 2009.

\bibitem{Vyas2007}
R. Vyas et al., ``Liquid crystal polymer (LCP): The ultimate solution for low-cost RF flexible electronics and antennas,'' \emph{Proc. IEEE Antennas Propag. Soc. Int. Symp.}, Jun. 2007, pp. 1729–1732

\bibitem{Mata2005}
A.~Mata, A.~J.~Fleischman, and S.~Roy, ``Characterization of Polydimethylsiloxane (PDMS) Properties for Biomedical
Micro/Nanosystems,'' \emph{Biomed. Microdevices}, 2005.

\bibitem{Wang2012}
Z. Wang, L. Zhang, Y. Bayram and J. L. Volakis, ``Embroidered Conductive Fibers on Polymer Composite for Conformal Antennas,'' \emph{IEEE Trans. Antennas Propag.}, vol. 60, no. 9, pp. 4141-4147, Sept. 2012.

\bibitem{AbelAEU}
A.~Zandamela, K. Schraml, S. Chalermwisutkul, D. Heberling, and Adam Narbudowicz,~``Digital pattern synthesis with a compact MIMO antenna of half-wavelength diameter,''~\emph{AEU – International Journal of Electronics and Communications},~vol.~135,~June~2021.

\bibitem{EuCAP2022}
A. Zandamela, N. Marchetti and A. Narbudowicz, ``Directional Modulation from a Wrist-Wearable Compact Antenna,'' \emph{2022 16th European Conference on Antennas and Propagation (EuCAP)}, Madrid, Spain, 2022, pp. 1-5.

\end{thebibliography}
\end{document}